\documentclass[aps,pra,twocolumn,superscriptaddress,noeprint,longbibliography]{revtex4-1}
\usepackage{bbm}
\usepackage{hyperref}
\usepackage{natbib}
\usepackage{amsmath}
\usepackage{verbatim}
\usepackage{color}
\usepackage{psfrag}
\usepackage{amsfonts}
\usepackage{amssymb}
\usepackage{amsbsy}
\usepackage{latexsym}
\usepackage{epsfig}
\usepackage{graphicx}
\usepackage{bm}
\usepackage{subfigure}
\usepackage{mathtools}

\begin{document}
\title{Dynamics of Topological Defects in a Rashba Spin-Orbit Coupled Bose-Einstein Condensate}
\author{Sheng Liu}\email{shengliu@ustc.edu.cn}
\author{Yong-Sheng Zhang}\email{yshzhang@ustc.edu.cn}
\affiliation{Key Laboratory of Quantum Information, University of Science and Technology of China, Hefei, 230026, China}
\affiliation{Synergetic Innovation Center of Quantum Information and Quantum Physics, University of Science and Technology of China, Hefei, 230026, China}
\affiliation{Hefei National Laboratory, University of Science and Technology of China, Hefei 230088, China}
\date{\today}

\begin{abstract}
We investigate the quench dynamics of a two-dimensional Rashba spin-orbit coupled Bose-Einstein condensate. Our study focuses on quenching the system from a zero-momentum phase to a plane-wave phase. During this quench, topological defects emerge in the form of vortices. These vortices and anti-vortices exhibit a random spatial distribution with equal numbers, mirroring the core principles of Kosterlitz-Thouless physics. In a uniform system, we observe an exponential scaling of both the vortex production time and the vortex number with the quench rate, consistent with the conventional Kibble-Zurek mechanism. The decay of which adheres to a logarithmic law, aligning with experimental observations.
\end{abstract}
\maketitle
\section{Intruduction}
In his seminal 1976 paper~\cite{Kibble1976}, Kibble examined the cooling process of the early universe and introduced the concept of cosmic strings, a type of topological defect, arising from symmetry breaking during this cooling. These cosmic strings are linear topological defects with potentially significant cosmological implications. Kibble's work laid the groundwork for understanding the formation of such structures in the early universe and their potential role in the formation of large-scale cosmic structures. Later, Zurek extended the concept of symmetry breaking-induced defects into the realm of condensed matter physics, as detailed in his works~\cite{Zurek1985,Zurek2005,Zurek2009}. This extension is now known as the Kibble-Zurek mechanism (KZM), which states that when a system undergoes a second-order phase transition at a finite rate, the emergence of topological defects is inevitable.

The dimensionality of the defects is determined by the specific system under consideration, while their size is governed by the rate at which the system traverses a second-order phase transition. Furthermore, both the size of the defects and the timing of their emergence exhibit scaling laws with respect to the rate of traversing the second-order phase transition. The morphology of these defects varies across different systems. For example, in a one-dimensional Ising model, defects manifest as domain walls, whereas in a ring-shaped supercooled helium system, they appear as line segments. The KZM has been the subject of numerous theoretical and experimental investigations across a wide array of systems~\cite{Damski2005,Damski2006,Damski2010,Dziarmaga2005,Lee2004,Lee2009,Lamacraft2007a,Saito2007a,Wu2017,Ye2018,Anquez2016,Uhlmann2007,Chuang1991b,Weiler2008,Clark2016,Labeyrie2016,Xu2014,Pyka2013}. Typically, the KZM is applied to conventional second-order phase transitions. Recently, the KZM has been successfully applied to first-order phase transitions~\cite{Duan2020}, yielding results that are in good agreement with theoretical predictions. This application of KZM to first-order transitions broadens the scope of the mechanism beyond its traditional domain of second-order phase transitions.

Phase transitions are typically orchestrated by manipulating system parameters, such as temperature or pressure. However, certain phase transitions are governed by intrinsic dynamical parameters. A notable example is the phase transition in Raman-type spin-orbit coupled Bose-Einstein condensates (SOCBEC), as referenced in the works by~\cite{Liu2019,Yi2020}, which is controlled by adjusting the Rabi coupling strength. This type of phase transition offers a unique avenue for exploring the interplay between dynamics and symmetry breaking. In this investigation, we focus on a two-dimensional Rashba SOCBEC, a system that has been successfully realized experimentally~\cite{Wu2016}. The system we examine here features Rashba SOC and a significant trapping potential in $z-$direction, and we employ a two-dimensional Gross-Pitaevskii equation (GPE), which is appropriate based on the aforementioned considerations. 

In the context of Rashba SOCBEC, vortex and anti-vortex pairs can be deliberately generated and manipulated by precisely designing and controlling specific system parameters. This mechanism of vortex and anti-vortex generation is analogous to that found in Kosterlitz–Thouless (KT) physics~\cite{Kosterlitz_1973}.  In this study, we apply the KZM to a first-order phase transition. The system generates an equal number of vortex and anti-vortex. In a uniform system, the emergence times and the numbers of the vortex adhere to scaling laws that are contingent upon the rates of the quenching parameter, as predicted by KZM.

\section{System Hamiltonian}
We examine a uniform two-dimensional Rashba SOCBEC~\cite{Ho2011,lin2011,Yun2012,Zhang2012,ZhangY2013}, which is characterized by a two-component, dimensionless GPE
\begin{equation}
i\partial_{t}\Phi(\bm r, t)=(H-\mu)\Phi(\bm r,t),
\end{equation}
where $\Phi(\bm r,t)=[\phi_1(\bm r,t),\phi_2(\bm r,t)]^{\mathrm{T}}$, $\bm r=(x,y)$, $\mu$ is the chemical potential which is introduced to control the total number of particles, and $H$ contains the single-particle part $H_0$ and the interaction-part $H_I$, which can be expressed as following
\begin{equation}
H_0=
\begin{pmatrix}
\frac{k_{\perp}^2}{2}+\gamma k_x+V(\bm r) & \gamma k_y-i\Omega\\
\gamma k_y+i\Omega & \frac{k_{\perp}^2}{2}-\gamma k_x+V(\bm r)
\end{pmatrix},
\end{equation}
and
\begin{equation}
H_I=
\begin{pmatrix}
g_{11}|\phi_1|^2+g_{12}|\phi_2|^2 & 0\\
0 & g_{21}|\phi_1|^2+g_{22}|\phi_2|^2
\end{pmatrix},
\end{equation}
where $k_{\perp}^2=k_x^2+k_y^2$, $\gamma$ is the spin-orbit coupling strength, $V(\bm r)$ is the trapping potential, and $\Omega$ is the Rabi coupling strength.

\begin{figure}
\centering
\includegraphics[width=8cm,height=3cm]{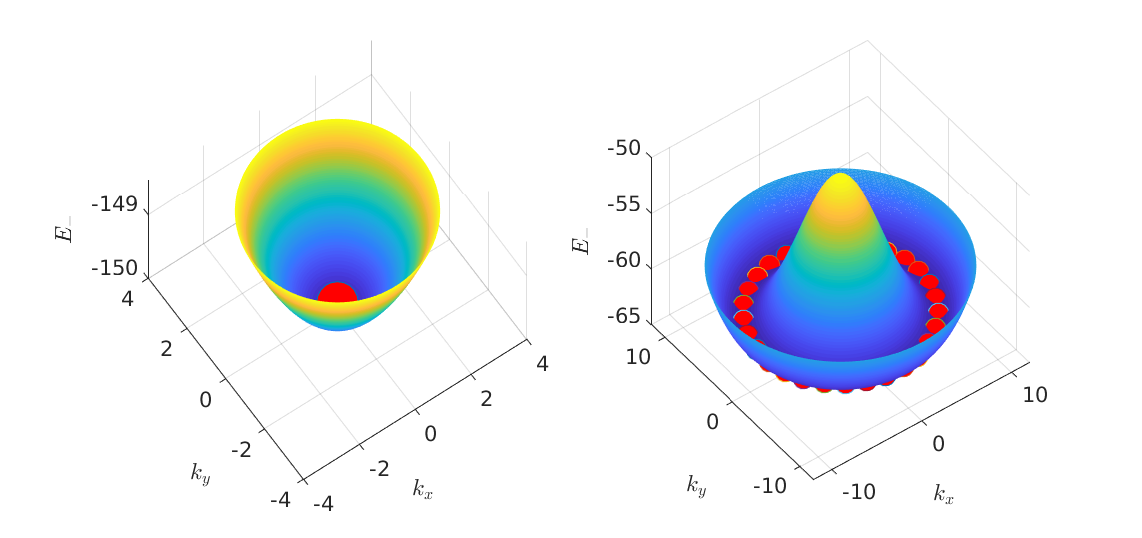}
\caption{The lower energy spectrum $E_{-}$ for the single-particle Hamiltonian in the absence of a trapping potential, left panel: zero-momentum phase ($\Omega>\Omega_{\mathrm{c}}$), right panel: plane-wave phase ($\Omega<\Omega_{\mathrm{c}}$). In both panels, the red-solid balls denote the locations of the lowest energy.}
\label{fig:spectrum}
\end{figure}

The Rashba-type SOC part can be expressed as $\gamma  k_x\sigma_z+\gamma k_y\sigma_x+\Omega\sigma_y$. The energy spectrum for the single-particle Hamiltonian without trapping potential is $E_{\pm}=\frac{ k_{\perp}^2}{2}\pm \sqrt{\gamma^2 k_{\perp}^2+\Omega^2}$, and the lower energy spectrum $E_{-}$ is plotted in Fig.~\ref{fig:spectrum}. Without trapping potential and interaction, the system condenses into $\bm k=0$ for $\Omega>\Omega_{\mathrm{c}}\equiv \gamma^2$, which we call the zero-momentum phase (ZMP, as indicated by the red-solid ball in the left panel of Fig.~\ref{fig:spectrum}). The system condenses into a superposition state containing different $\bm k=(k_x,k_y)$, where $\bm k$ lies on the ring $k_x^2+k_y^2=\gamma-\frac{\Omega^2}{\gamma^2}\equiv k_c^2$ (red-solid balls in the right panel of Fig.~\ref{fig:spectrum}) for $\Omega<\Omega_{\mathrm{c}}$. For the nonzero momentum phase, this degeneracy can be lifted by interaction and trapping potential, and the system will choose only two opposite momenta $\bm k_0$ and $-\bm k_0$ from the degenerate ring~\cite{Wu_2011,Zhou_2013}. If we tune the parameter $\Omega$ from $\Omega_i>\Omega_c$ to $\Omega_f<\Omega_c$ with a finite rate, then each point will choose its own $\bm k_0$. Due to the rotational symmetry of the Hamiltonian, there will be vortex excitations. The term $\Omega\sigma_y$ in the Hamiltonian indicates that the quench process will not introduce additional angular momentum into the system. The numerical simulation confirms that the number of vortices and anti-vortices are equal.

\section{Numerical Simulation}
The original trapping potential can be described as $V_{\rm 3D}(\bm r,z)=V(\bm r)+\frac{1}{2}m\omega_z^2 z^2$ where the excitations in $z-$direction are curtailed by selecting a large confinement in this direction, thereby creating a quasi-two-dimensional system. 

To simulate the process of vortex production, it is essential to initially seed the wavefunction with quantum noise. As detailed in~\cite{Blakie2008}, we can express the quantum noise in terms of the eigenmodes of the Bogoliubov-de Gennes (BdG) equations~\cite{BdG}, with the expansion coefficients being randomly selected from a Gaussian distribution~\cite{QuantumNoise}. However, solving the BdG equations for a two-dimensional system presents a challenge. Therefore, we opt for a Gaussian distribution noise and incorporate it into the ground-state wavefunction as follows:
\begin{equation}
\Phi(\bm r,t=0) = \Phi_0(\bm r)+\varepsilon[\eta_1(\bm r)+i\eta_2(\bm r)],
\end{equation}
where $\Phi_0(\bm r)$ is the ground-state wavefunction for the system, and $\{\eta_1(\bm r),\eta_2(\bm r)\}$ are the quantum fluctuation which satisfies Gaussian distribution $\sim N(0,1)$ at each point $\bm r$ and $\varepsilon\ll 1$ controls the level of noise. 
This approximation has been employed in several studies~\cite{Symes2018} and has demonstrated excellent agreement with experimental outcomes~\cite{Liu2019}. Furthermore, to mitigate high-energy collective excitations, we incorporate a small positive dissipation parameter $\beta \ll 1$ into the GPE~\cite{Choi1998,pitaevskii1959,Gauthier1264}:
\begin{equation}
(i-\beta)\partial_t\Phi(\bm r,t)=(H-\mu)\Phi(\bm r,t).
\end{equation}

To mitigate the effects of inhomogeneity in our numerical simulation, we have implemented a cylinder-symmetrical box potential with a smooth boundary, expressed as $V(\bm r)= \frac{V_0}{2}[\tanh(|\mathbf{r}|-R)+1]$, where $R$ and $V_0$ represent the radius and height of the box potential, respectively. We use package XMDS2~\cite{Dennis2013} for all the simulations.


\section{Kibble-Zurek Mechanism}
In the case of a homogeneous system undergoing a second-order phase transition, we introduce a parameter $\epsilon$ to quantify the deviation from the critical point. For instance, $\epsilon$ could represent thermodynamic parameters such as temperature (e.g.,~\cite{Zurek2009}) or control parameters of the Hamiltonian (e.g.,~\cite{Clark2016}). As the critical point is approached, the correlation length $\xi$ and relaxation time $\tau$ diverge according to the following expressions:
\begin{equation}
\xi=\xi_0/|\epsilon|^\nu,\quad
\tau=\tau_0/|\epsilon|^{\nu z},
\end{equation}
where $\xi_0$ and $\tau_0$ are constants specific to the system, and the critical exponents $\nu$ and $z$ are characteristic of the universality class of the phase transition.

In accordance with KZM, for a linear quench described by $\epsilon=\frac{\Omega(t)-\Omega_c}{\Omega_c}$, where $\tau_q$ is the quench time, and $\Omega(t) = \frac{\Omega_f-\Omega_i}{\tau_q}t+\Omega_i$, we derive that $\hat t-t_C=(\tau_0\tau_q^{\nu z})^{\frac{1}{1+\nu z}}\sim \tau_q^{\frac{\nu z}{1+\nu z}}$, where $t_C$ indicates the time when $\Omega(t_c)=\Omega_c$. The correlation length that freezes in at $\hat{t}$ is given by
\begin{equation}
\hat \xi=\xi_0({\tau_q}/{\tau_0})^{\frac{\nu}{1+\nu z}},
\end{equation}
the scaling law of defects number in two-dimensional system is given by $N_v\sim \tau_q^{-\frac{2\nu}{1+\nu z}}$

\section{Scaling laws}
\begin{figure}
\centering
\includegraphics[width=8.5cm,height=4.5cm]{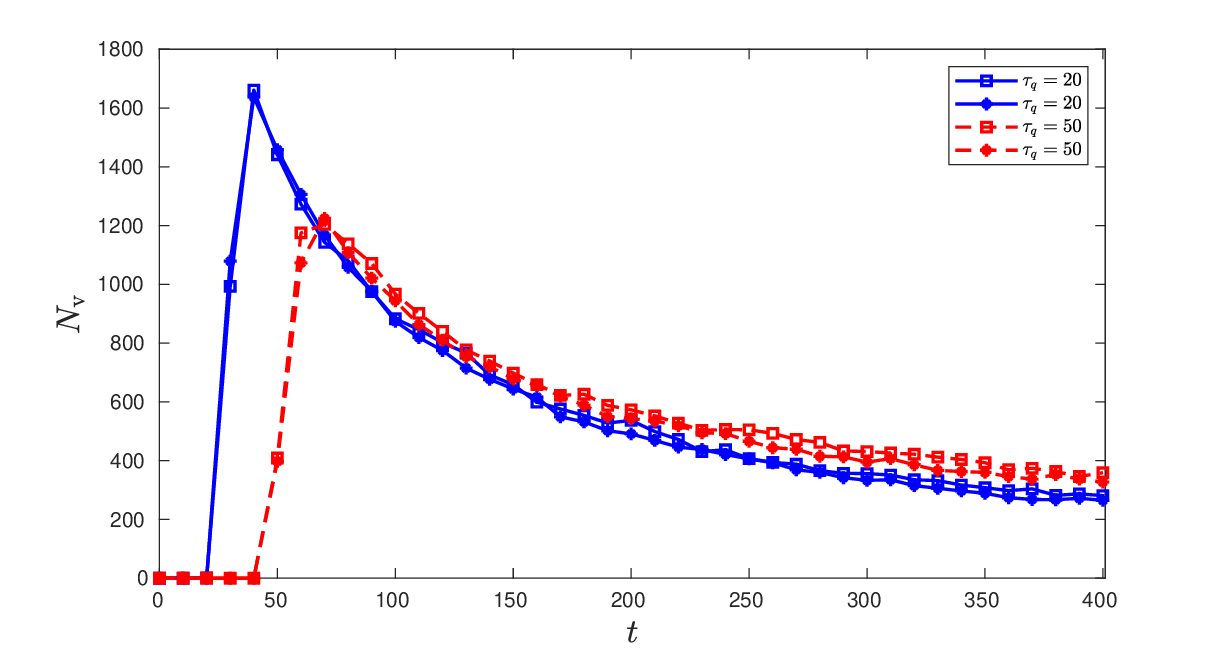}
\caption{The time-dependent evolution of vortex and anti-vortex numbers for two different quench times: $\tau_q=20$ and $\tau_q=50$. In the plots, blue squares and blue stars represent vortices, while red squares and red stars represent anti-vortices, respectively.}
\label{fig:vortex_number}
\end{figure}

In Fig.~\ref{fig:vortex_number}, we depict the evolution of the vortex and anti-vortex numbers over time for quench times $\tau_q=20$ and $\tau_q=50$. Initially, the system is nearly devoid of vortices or anti-vortices. Once $\Omega$ surpasses the critical point, the formation of vortices and anti-vortices commences. By the conclusion of the time evolution, the system exhibits approximately an equal number of vortices and anti-vortices. This equilibrium also suggests that the mechanism of vortex formation is analogous to that observed in KT physics.

\begin{figure}
\centering
\includegraphics[width=8.5cm,height=4.5cm]{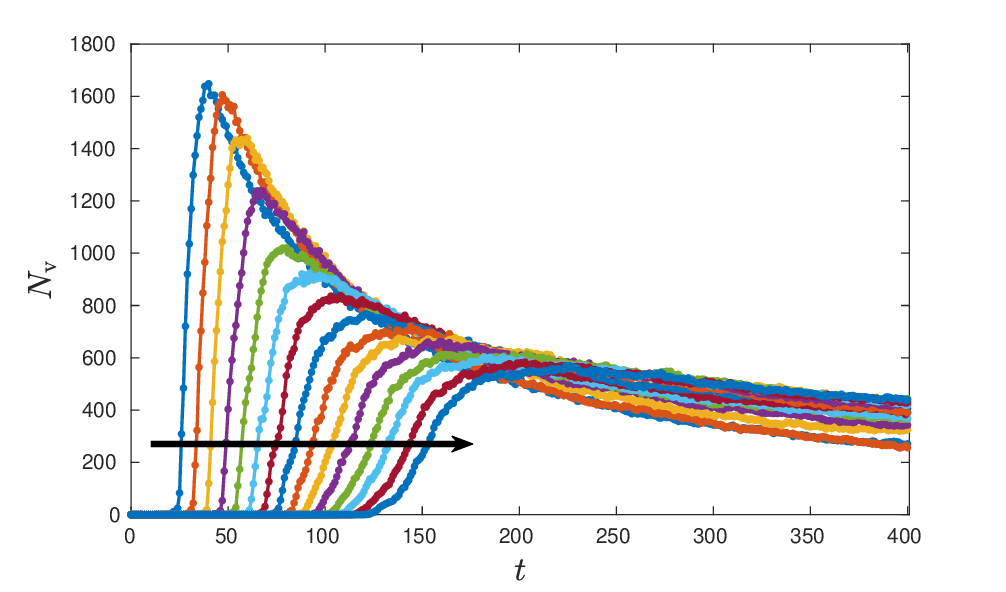}
\caption{The number of vortices as a function of time for various quench times (from $\Omega_i=2\Omega_c$ to $\Omega_f=0.1\Omega_c$), ranging from $\tau_q=20$ to $\tau_q=160$, as indicated by the black arrow.}
\label{fig:vortex_time}
\end{figure}

The vortex numbers as a function of time for different quench times are illustrated in Fig.~\ref{fig:vortex_time}. For each quench time $\tau_q$, the critical $\Omega_c$ is attained at time $t_{\mathrm{C}}=\frac{\Omega_i-\Omega_c}{\Omega_i-\Omega_f}\tau_q$. As predicted by KZM, vortex formation occurs after the time $t_{\mathrm{C}}$, as depicted in Fig.~\ref{fig:vortex_time}. To determine the time of vortex formation, we establish a threshold vortex count of 5.

\begin{figure}
\centering
\includegraphics[width=8.5cm,height=4cm]{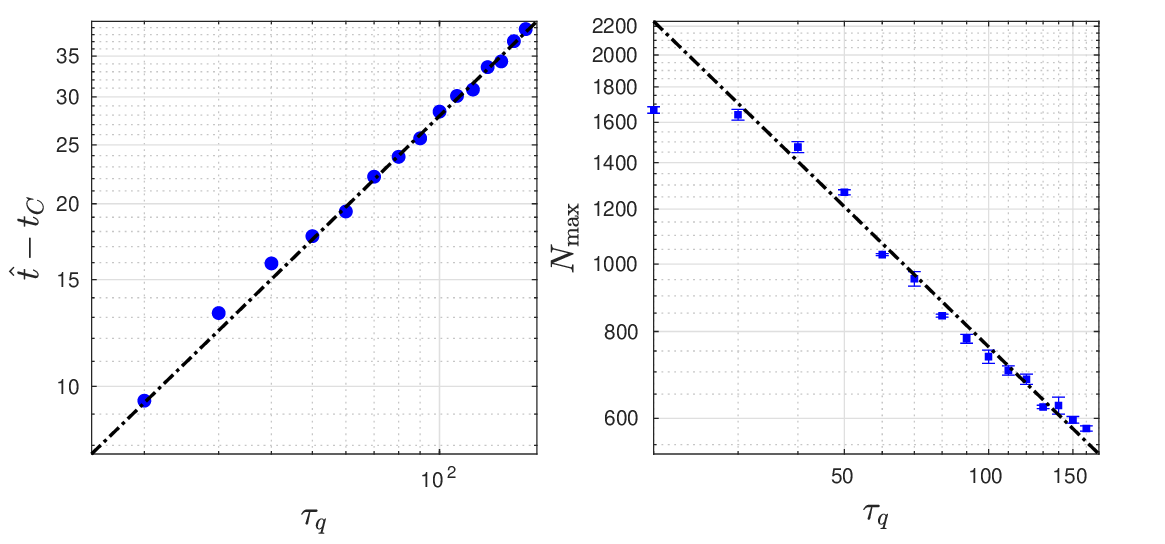}
\caption{The delay in vortex formation times and the corresponding vortex numbers as functions of quench times. Blue circles denote the numerical data points, while the black dash-dotted lines represent the fitted slopes, which are approximately $0.68\pm 0.03$ and $-0.67\pm 0.06$, respectively.}
\label{fig:scaling_law}
\end{figure}

In Fig.~\ref{fig:scaling_law} (left), we present $\hat{t}-t_{\mathrm{C}}$ as a function of quench times. By fitting the data points, we derive the scaling law
\begin{equation}
\hat{t}-t_{\mathrm{C}}\sim \tau_q^{0.68\pm 0.03},
\label{eq:t_law}
\end{equation}
this gives us $\frac{\nu z}{1+\nu z}=0.68\pm 0.03$, and this scaling exponent closely resembles that found in the inhomogeneous KZM scenario. 

To determine the dynamical critical exponent $z$, we must ascertain the vortex numbers as a function of quench times. The correlation length is bounded by the healing length, which is a finite value within a BEC system. This limitation places an upper bound on the maximum vortex number that can be generated during the quench process. Furthermore, vortex and anti-vortex pairs will annihilate upon colliding after the quench process. As observed in Fig.~\ref{fig:vortex_time} and Fig.~\ref{fig:scaling_law}, the vortex numbers nearly saturate for very rapid quench times and asymptotically approach a constant value for very slow quench times.

In Fig.~\ref{fig:scaling_law} (right), we present the maximum vortex number after the quench process $N_{\rm max}$ as function of quench times. The data points corresponding to quench times $\tau_q=20,30$ are not used for fitting since the vortex numbers are saturated due to the very fast quench. The fitting gives
\begin{equation}
N_{\rm max}\sim \tau_q^{-0.67\pm 0.06},
\label{eq:N_law}
\end{equation}
 and we have $\frac{2\nu}{1+\nu z}=0.67\pm 0.06$ which agrees very well with the experiment conducted in a two-dimensional superfluid~\cite{Lee2024}. Combine with Eq.~(\ref{eq:t_law}) and Eq.~(\ref{eq:N_law}), we can extract the scaling exponents $\nu=1.03$ and $z=2.02$. The Rashba spin-orbit coupling affects the static critical exponent $\nu$, which is $\nu=1.03$ in our case and $\nu=\frac{1}{2}$ for KZM applied to a second order phase transition within mean-field framework. The dynamical critical exponent $z=2$ results from the quartic kinetic energy $\sim k^4$ near the critical phase transition point~\cite{Clark2016,Wu2017}.

\section{Vortex decay dynamics}
\begin{figure}
\centering
\includegraphics[width=8.5cm,height=4cm]{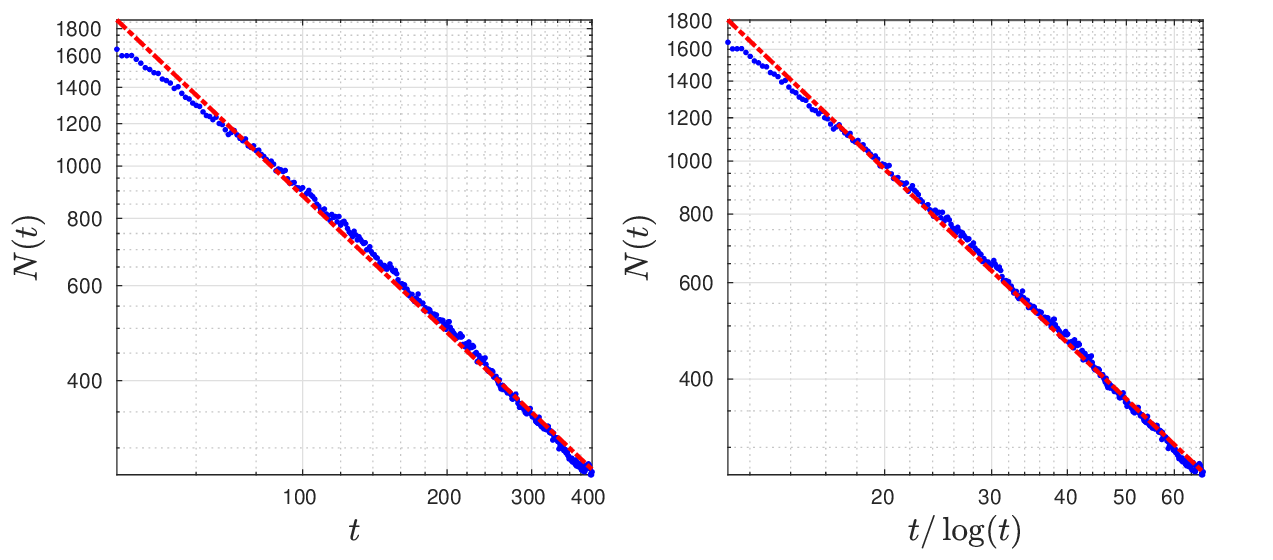}
\caption{The decay dynamics of vortices for a quench time of $\tau_q=20$ are depicted in the two figures. In each, blue circles represent the data obtained from numerical simulations, and the red dashed-dotted lines represent the corresponding fits. On the left, the vortex decay is plotted against time $t$, with the fit yielding $N(t)\sim t^{-0.84\pm 0.01}$. On the right, the vortex decay is plotted against $t/\log(t)$, with the fit yielding $N(t)\sim (t/\log(t))^{-1.06\pm 0.01}$.}
\label{fig:vortex_decay}
\end{figure}

Following the quench process, the system transitions into a state of non-equilibrium dynamics, during which vortices and anti-vortices begin to form, as depicted in Fig.~\ref{fig:vortex_time}. The number of vortices swiftly reaches a peak and subsequently enters a decay phase. If we disregard the spatial structure of the vortices, their interactions can be approximated as point interactions. In this simplified scenario, a mean-field description can be succinctly expressed as $\frac{dN(t)}{dt}=-\kappa N^2(t)$, This leads to the solution $N(t)\sim \frac{1}{t}$. However, when the spatial structure of the vortices is taken into account, the decay relationship is refined with a logarithmic correction~\cite{Bray1995,Bray2000}
\begin{equation}
N(t)\sim\frac{1}{t\log(t)},
\end{equation}
As $t$ becomes significantly larger than one, this logarithmic relationship tends towards an asymptotic behavior described by $N(t)\sim t^{-0.85}$. In Fig.~\ref{fig:vortex_decay}, we present two fitting protocols that capture this decay behavior. The left-hand plot demonstrates an excellent fit for $N(t)\sim t^{-0.84}$, while the right-hand plot provides a fit for $N(t)\sim (t/\log(t))^{-1.06}$. The $\log(t)$ part arises from the effective Coulomb interaction $\frac{1}{r}$ between two vortices of distance $r$~\cite{Jelic_2011,Yurke1993}. Both fitting protocols are in close alignment with the theoretical predictions, underscoring the validity of the logarithmic decay model in describing the vortex dynamics post-quench.

\begin{figure}
\centering
\includegraphics[scale=0.5]{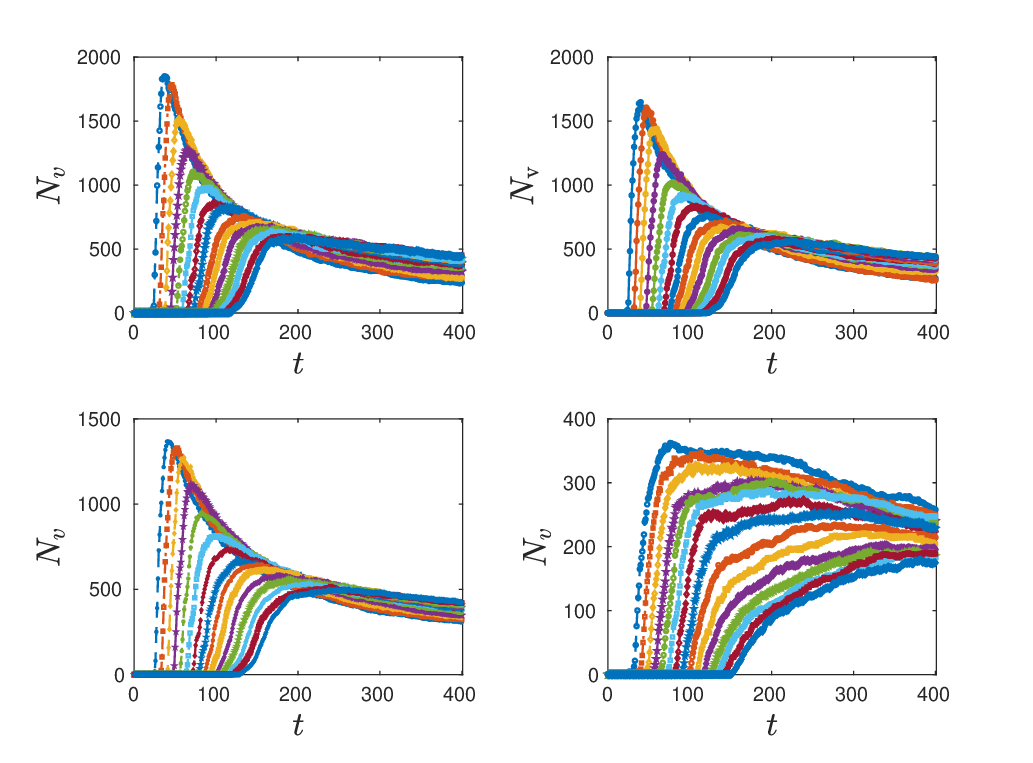}
\caption{Vortex numbers with time for different quench times. (a): from $\Omega_i=2\Omega_c$ to $\Omega_f=0$, (b): from $\Omega_i=2\Omega_c$ to $\Omega_f=0.1\Omega_c$, (c): from $\Omega_i=2\Omega_c$ to $\Omega_f=0.2\Omega_c$, (d): from $\Omega_i=2\Omega_c$ to $\Omega_f=0.5\Omega_c$,}
\label{fig:vortex_time_quenches}
\end{figure}

In Fig.~\ref{fig:vortex_time_quenches}, we have displayed the outcomes for quench protocols with final frequencies $\Omega_f=0$, $\Omega_f=0.1\Omega_c$, $\Omega_f=0.2\Omega_c$, and $\Omega_f=0.5\Omega_c$. Specifically, for the quench protocol with $\Omega_f=0.5\Omega_c$, the number of vortices rapidly reaches saturation for quench times exceeding 50, as this condition approaches the adiabatic regime.

\begin{figure}
\centering
\includegraphics[scale=0.4]{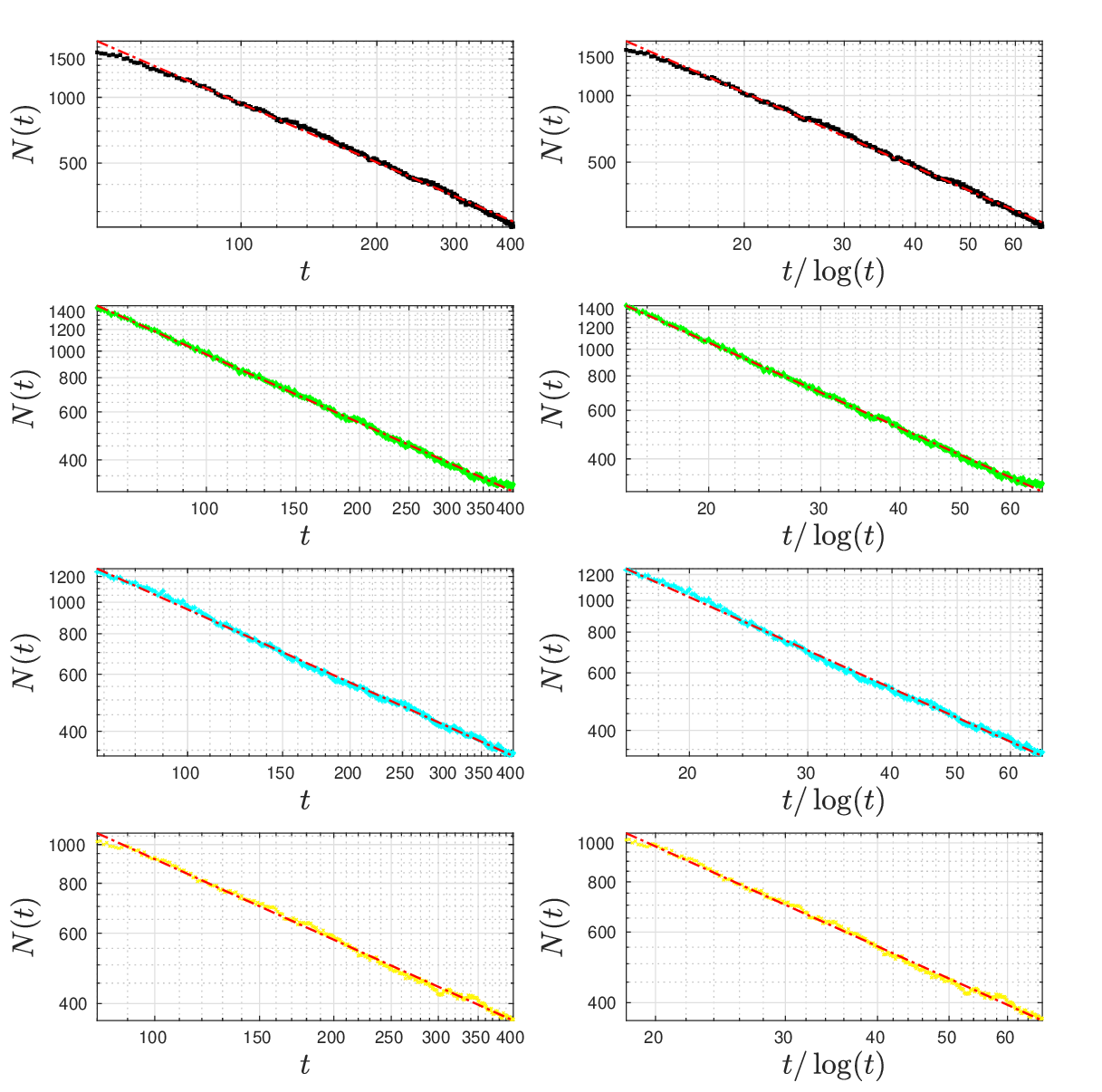}
\caption{Vortex decaying dynamics with quench from $\Omega_i=2\Omega_c$ to $\Omega_f=0.1\Omega_c$ for quench time $\tau_q=30,40,50,60$ (from up to bottom). The colored discrete data points are numerical simulations, and the red dash-dotted lines are the fitting lines. Left columns show the fittings for $t$, the slopes are $\{-0.90,-0.83,-0.75,-0.67\}$. Right columns show the fittings for $t/\log(t)$, the slopes are $\{-1.11,-1.03,-0.93,-0.83\}$}
\label{fig:vortex_decay_tq_3456}
\end{figure}

In our exploration of how vortex decay dynamics vary with different quench times, we have presented the evolution of vortex numbers over time in Fig.~\ref{fig:vortex_decay_tq_3456}. The figure reveals a distinct pattern: for both very large and very small quench times, the decay curves of the vortices significantly diverge from the theoretical decay rates of $\frac{1}{t}$ and $\frac{\log(t)}{t}$. In contrast, for intermediate quench times, particularly around 40, the decay curves are in good agreement with the experimental data.

\begin{figure}
\centering
\includegraphics[width=8.5cm,height=4cm]{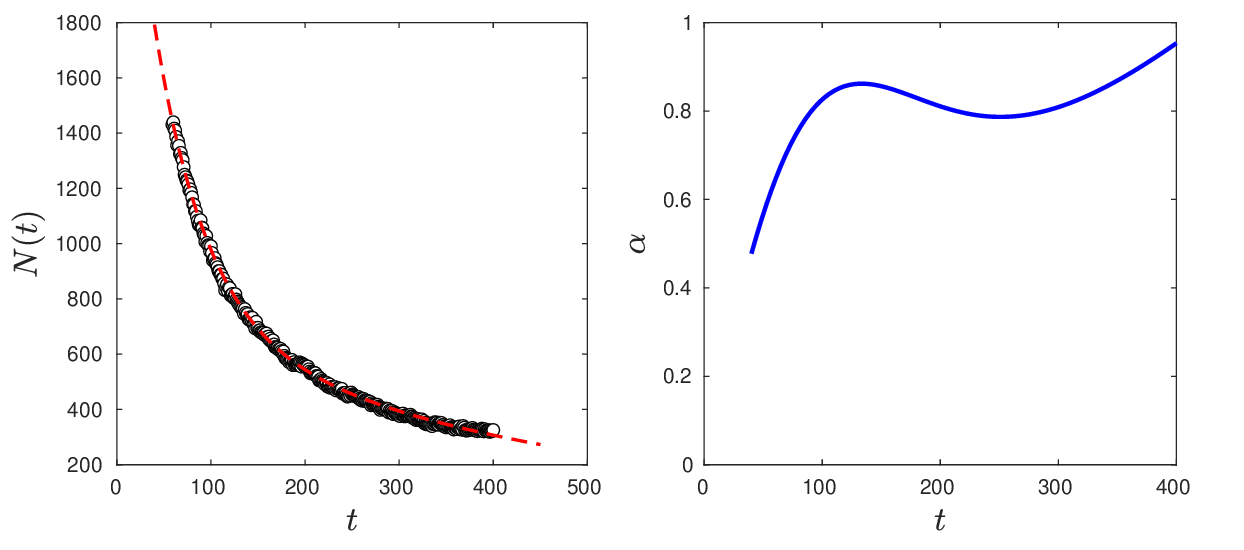}
\caption{The vortex number as a function of time, following a quench from $\Omega_i = 2\Omega_c$ to $\Omega_f = 0.1\Omega_c$ with a quench time $\tau_q = 40$, is illustrated. On the left panel, the black circles represent the numerical data points, while the red dashed line corresponds to the fitting using the form $a e^{bt} + ce^{dt}$ (coefficient of determination $R^2=0.9989$). The right panel displays the calculated time-dependent decay exponent $\alpha$.}
\label{fig:decay_exponent}
\end{figure}

To enhance our understanding of the vortex decay dynamics, we have fitted the numerical data $N(t)$ with bi-exponential form $a e^{bt}+ce^{dt}$. Subsequently, we calculated the time-dependent decay exponent using the formula $\alpha = -\frac{d\log (N)}{d\log(t)}$~\cite{Yurke1993}. The results of this calculation are depicted in Fig.~\ref{fig:decay_exponent}. The data fit with the bi-exponential form aligns very well with the numerical data. Initially, the time-dependent decay exponent is relatively low, around $0.5$, and then it gradually increases to $0.85$. Over longer periods, the decay exponent tends to approach unity. These observations are in excellent agreement with the numerical simulations conducted in the 2D $XY$ model, as reported in~\cite{Yurke1993}.

\section{Spatial distribution of vortex}
In two-dimensional system, the interaction between point-like vortices has form
\begin{equation}
E=-C\Gamma_i\Gamma_j\log(|{\bm r}_i-{\bm r}_j|/\xi),
\end{equation}
where $C$ is a positive constant, $\Gamma_i,\Gamma_j$ are the circulation numbers of the vortices, $\bm r$ are the vortices positions, and $\xi$ is the healing length of the system.

\begin{figure}
\centering
\includegraphics[width=8.5cm,height=4cm]{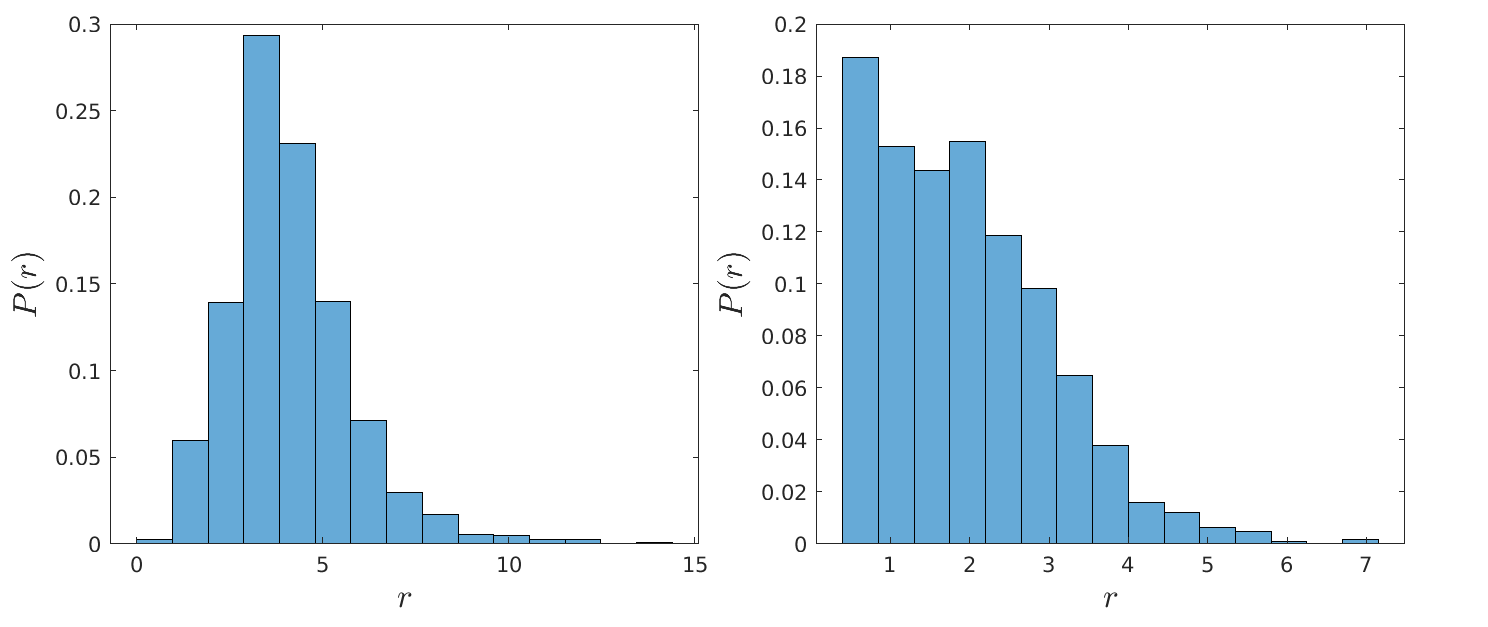}
\caption{Probability distribution of the vortex distance between positive-positive vortex (left) and between positive-negative vortex (right). The vortex profile is extracted at time $t = 80$ for quench time $\tau_q=20$ and $\Omega_f=0.1\Omega_c$.}
\label{fig:vortex_distribution}
\end{figure}

We track each vortex, pair it with its nearest-neighboring vortex, and record the distance between them. If two vortices possess circulation numbers of opposite signs, they are inclined to be closer in proximity to lower the total energy. Conversely, if they share the same circulation number, their energy is minimized when they are positioned further apart. As depicted in Fig.~\ref{fig:vortex_distribution}, we have illustrated the probability distribution of the distances between vortices with the same circulation number and those with opposite circulation numbers. This visualization clearly demonstrates that vortices with circulation numbers of opposite signs are more likely to cluster together to achieve a lower total energy state, whereas the opposite is true for vortices with circulation numbers of the same sign.

\section{Conclusion}
In conclusion, we have investigated the quench dynamics of Rashba SOCBEC in uniform trapped systems. Throughout the quenching process, an equal number of vortices and anti-vortices emerge. We applied the KZM to the vortex production process and discovered that both the vortex numbers and the times of vortex emergence adhere to scaling laws relative to the rate of the quench parameter. We demonstrate that the presence of Rashba spin-orbit coupling affects the scaling exponent of the emergence time of the topological defects, which is found to be $\frac{\nu z}{1+\nu z}=0.68\pm 0.03$ in our numerical simulations, as compared with the mean-field value $\frac{\nu z}{1+\nu z}=\frac{1}{2}$~\cite{Liu2019} obtained with KZM applied to a second order phase transition. However, the dynamical critical exponent $z=2.02$, as determined from our simulations, is in excellent agreement with experimental data~\cite{Lee2024} and numerical simulations across various systems~\cite{Wu2017}. We also have studied the vortex decaying dynamics after the quench process, the results we obtained here is consistent with experiment~\cite{Liu2021}, and the $\log(t)$ part which arising from the $\frac{1}{r}$ Coulomb interaction is identified in our numerical simulations. Furthermore, the distinctive feature of equal abundance of vortices and anti-vortices in the system under study renders it an ideal candidate for delving into Onsager's vortex physics and the realm of quantum turbulence~\cite{Onsager1949,Carlo2014,Gauthier1264}. 

This work is supported by the National Natural Science Foundation of China (Grant No. 92065113), the Innovation Program for Quantum Science and Technology (2021ZD0301201), the University Synergy Innovation Program of Anhui Province (Grant No. GXXT-2022-039) and the Fundamental Research Funds for the Central Universities (WK2030000088).

\appendix*
\section{}
\subsection{Dimensionless Gross-Pitaevskii Equation}
The system is described by the following two-component Gross-Pitaevskii equation (GPE)
\begin{equation}
i\hbar\partial_t\Phi(\bm r,t)=(H-\mu)\Phi(\bm r,t),
\end{equation}
where $\Phi(\bm r,t)=[\phi_1(\bm r,t),\phi_2(\bm r,t)]^{\mathrm{T}}$, $\bm r=(x,y)$. The system Hamiltonian $H$ contains a single-particle part $H_0$ and an interaction part $H_I$. For the single-particle part
\begin{equation}
H_0=
\begin{pmatrix}
\frac{\hbar^2 k_{\perp}^2}{2m}+\gamma \hbar k_x+V(\bm r) & \gamma\hbar k_y-i\hbar\Omega\\
\gamma \hbar k_y+i\hbar\Omega & \frac{\hbar^2 k_{\perp}^2}{2m}-\gamma \hbar k_x+V(\bm r)
\end{pmatrix},
\label{eq:single_particle}
\end{equation}
and for the interaction part
\begin{equation}
H_I=
\begin{pmatrix}
g_{11}|\phi_1|^2+g_{12}|\phi_2|^2 & 0\\
0 & g_{21}|\phi_1|^2+g_{22}|\phi_2|^2
\end{pmatrix},
\end{equation}
where $k_{\perp}^2=k_x^2+k_y^2$, $\gamma$ is the spin-orbit coupling strength, $V(\bm r)$ is the trapping potential, $\Omega$ is the Rabi coupling strength, $\{g_{ij}=\frac{4\pi\hbar^2 a_{ij}}{m a_z}\}$ are the reduced two-dimensional interaction strength where $\{a_{ij}\}$ are the $s-$wave scattering lengths and $a_z=\sqrt{\frac{h}{m\omega_z}}$.

We choose $t_0=1/\omega_z$, $a_0=\sqrt{\frac{\hbar}{m\omega_z}}$ and $E_0=\hbar \omega_z$ as the units of time, length and energy, respectively. The dimensionless GPE follows
\begin{equation}
i\partial_{t^{\prime}}\Phi^{\prime}(\bm r^{\prime}, t^{\prime})=(H_0^{\prime}+H_I^{\prime})\Phi(\bm r^{\prime},t^{\prime}),
\end{equation}
$H_0^{\prime}$ and $H_I^{\prime}$ can be expressed as following
\begin{equation}
H_0^{\prime}=
\begin{pmatrix}
\frac{k_{\perp}^{\prime 2}}{2}+\gamma^{\prime} k_x+V^{\prime}(\bm r^{\prime}) & \gamma^{\prime} k_y-i\Omega^{\prime}\\
\gamma^{\prime} k_y+i\Omega^{\prime} & \frac{k_{\perp}^{\prime 2}}{2}-\gamma^{\prime} k_x+V^{\prime}(\bm r^{\prime})
\end{pmatrix},
\end{equation}
and
\begin{equation}
H_I^{\prime}=
\begin{pmatrix}
g_{11}^{\prime}|\phi_1^{\prime}|^2+g_{12}^{\prime}|\phi_2^{\prime}|^2 & 0\\
0 & g_{21}^{\prime}|\phi_1^{\prime}|^2+g_{22}^{\prime}|\phi_2^{\prime}|^2
\end{pmatrix},
\end{equation}
where $\gamma^{\prime}=\frac{\gamma}{a_0\omega_z}$, $\Omega^{\prime}=\frac{\Omega}{\omega_z}$ and $g_{ij}^{\prime}=\frac{g_{ij}}{a_0^2E_0}$. All the variables with prime are dimensionless, hereafter will be omitted for simplicity.

\subsection{Numerical parameters}
\begin{figure}
\centering
\includegraphics[width=8cm,height=4cm]{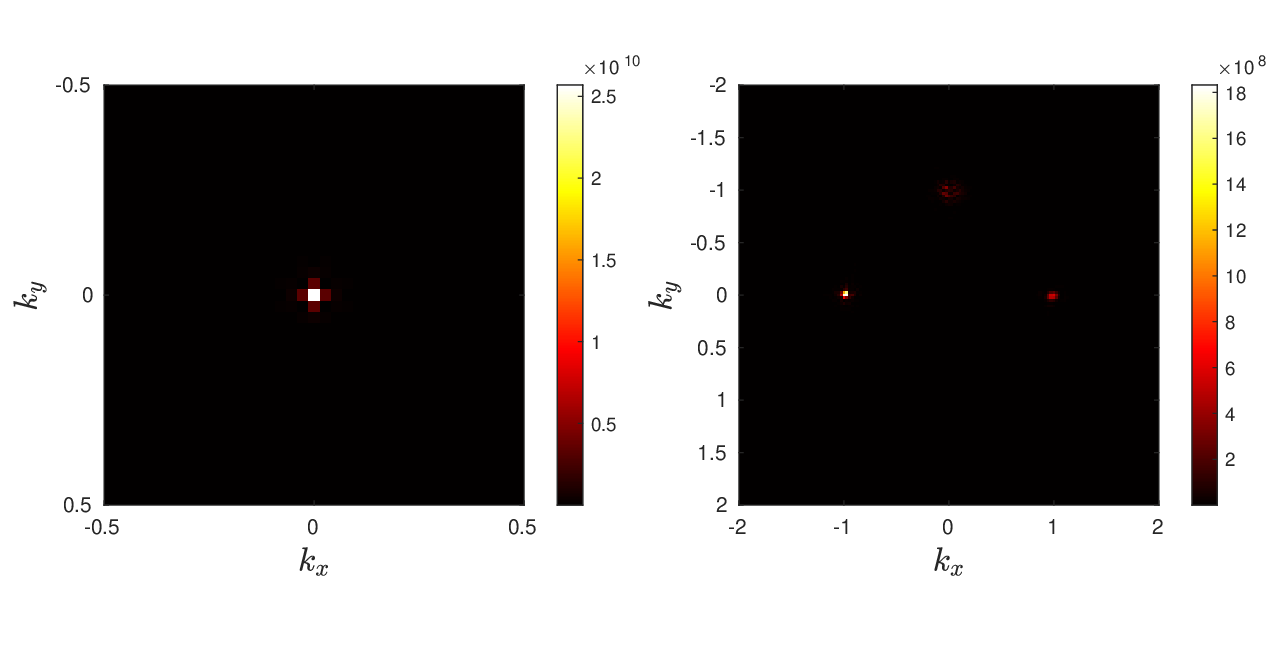}
\caption{Left: momentum distribution of groundstate in zero-momentum phase ($\Omega=2\Omega_c$), right: momentum distribution of groundstate in plane-wave phase ($\Omega=0.2\Omega_c$).}
\label{fig:GS_both}
\end{figure}

We set the interaction parameters to $g_{11}=g_{22}\equiv g$ and $g_{12}=g_{21}=0.8g$ ensuring the miscibility condition $g_{11}g_{22}-g_{12}^2>0$ is satisfied. We select $\gamma=1$ based on the experiment~\cite{Wu2016} which gives $\Omega_c=1$. The chemical potential is set to $\mu=1$ which gives the initial total particle number $N=\int [|\phi_1(\bm r,t)|^2+|\phi_2(\bm r, t)|^2]d^2\bm r\approx 1.03\times 10^5$. The initial Rabi frequency $\Omega_{i}=2\Omega_{c}$ and the final Rabi frequencies are $\Omega_{f}\in \{0,0.1\Omega_c,0.2\Omega_c,0.5\Omega_c\}$. The dissipation parameter is set to $\beta=10^{-3}$ which gives the total particle number $N\approx 6.2\times 10^{4}$ after the quench process. For each quench time $\tau_q$, we conduct the simulation $N=10$ times, each with distinct initial random-seeded wavefunctions. In Fig.~\ref{fig:GS_both}, we have depicted the ground states for the zero-momentum phase (left panel) and the plane-wave phase (right panel). It is evident that the zero-momentum phase exhibits a single momentum state of zero for the BEC, while the plane-wave phase features the BEC with opposite momenta.

\begin{figure}
\centering
\includegraphics[width=9cm,height=3.8cm]{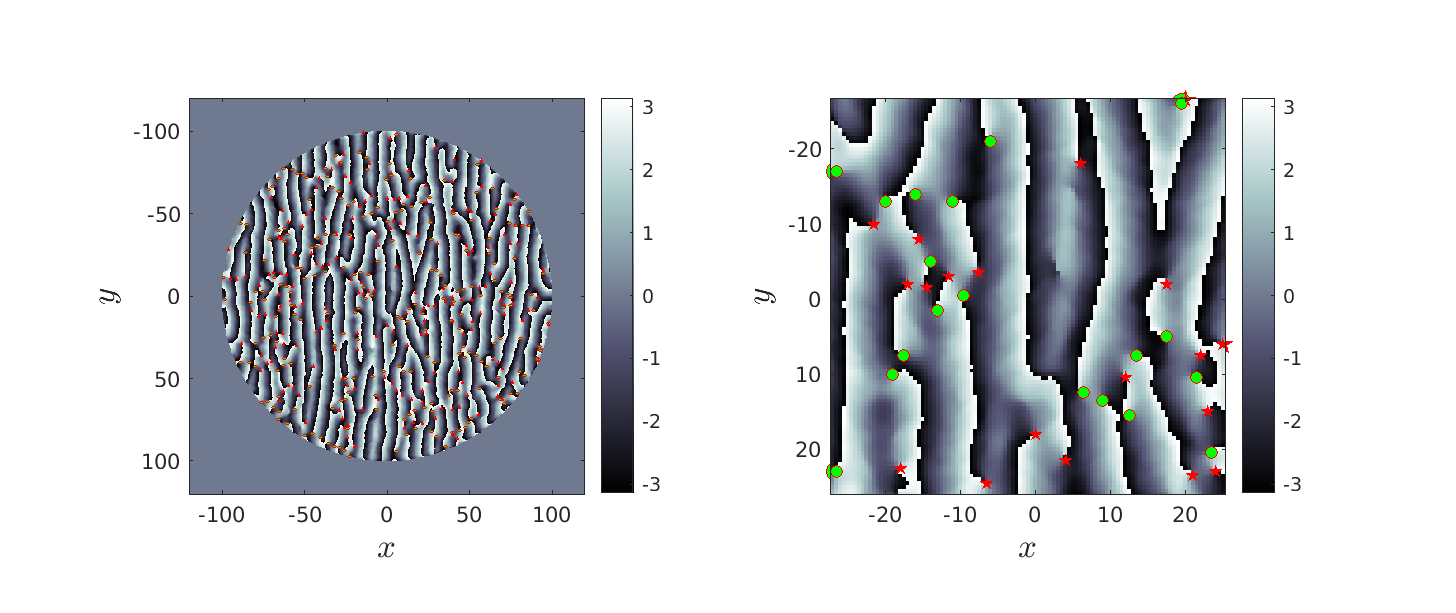}
\caption{Left panel: vortex distribution in the phase profile. Right panel: Enlargement for selected area. The locations of vortices and anti-vortices are marked by red pentagons and green circles, respectively.}
\label{fig:domain_demo}
\end{figure}

In Fig.~\ref{fig:domain_demo}, we depict the distribution of vortices in both the density and phase profiles. The positions of the vortices are identified using the phase profile of the wave function. The phase plot reveals that each open domain line terminates with one vortex and one anti-vortex, whereas closed domain rings contain no vortices. Throughout the quench dynamics, domain rings are initially formed, which then break apart into domain lines, simultaneously generating one vortex and one anti-vortex at the ends of these lines. Post-quench, vortices and anti-vortices can coalesce, regardless of whether they emerged from the same domain ring.


%

\end{document}